\documentclass[12pt,a4paper]{article}

\usepackage{graphics,epsfig}

\begin{document}

\newcommand{\goo}{\,\raisebox{-.5ex}{$\stackrel{>}{\scriptstyle\sim}$}\,}

\newcommand{\loo}{\,\raisebox{-.5ex}{$\stackrel{<}{\scriptstyle\sim}$}\,}

\begin{center}

{\large \bf On possibility of synthesizing superheavy elements in nuclear 
explosions}

\end{center}

\vspace{0.5cm}

\begin{center}

{\Large Alexander~Botvina$^{a,b}$, Igor~Mishustin$^{a,c}$, 
Valery~Zagrebaev$^{a,d}$ and Walter~Greiner$^{a}$}\\

\end{center}

\begin{center}

{\it

$^a$Frankfurt Institute for Advanced Studies, J.W. Goethe
University, D-60438 Frankfurt am Main, Germany\\

$^b$Institute for Nuclear Research, Russian Academy of Sciences,
117312 Moscow, Russia\\

$^c$Kurchatov Institute, Russian Research Center, 123182 Moscow,
Russia\\ 

$^d$Joint Institute for Nuclear Research, 141980 Dubna, Russia\\ }

\end{center}

\normalsize


\begin{abstract}

The possibility to produce superheavy elements in the course of 
low-yield nuclear explosions is analyzed within a simple kinetic model 
which includes neutron capture, $\gamma$-emission, fission and particle
evaporation from excited nuclei. We have calculated average numbers of 
absorbed neutrons as well as mass distributions of U and Cm nuclei exposed 
to an impulsive neutron flux as functions of its duration. It is demonstrated 
that detectable amounts of heavy nuclei absorbing from 20 to 60 neutrons 
may be produced in this process. According to an optimistic scenario, after 
multiple $\beta$-decay such nuclei may reach the long-living elements of the 
predicted ``island of stability''. 

\end{abstract}

\vspace{0.2cm}


\vspace{0.5cm}

{\bf 1. Introduction}

\vspace{0.5cm}

Synthesis of new  elements is one of the main goals of nuclear physics 
during the last century. First elements heavier than uranium were 
synthesized in nuclear reactors via neutron-capture reactions followed by 
$\beta$-decay. These reactions were efficient enough to produce 
elements up to Z=100 (Fm). Nuclei with larger Z are produced with 
accelerator-based experiments, most often via the fusion reaction 
involving $\alpha$-particles and heavy ions. 
In heavy-ion induced fusion reactions new nuclei up to Z=118 were
synthesized during the last decades \cite{review1, review2, review3}.
However, the nuclei produced in the cold fusion reactions \cite{review1,
review2} are situated just along the proton drip-line, thus being very
neutron-deficient and short-lived. Fusion of actinides with $^{48}$Ca
leads to more neutron-rich superheavy nuclei with longer half-lives
\cite{review3}. However, they are still far from the predicted "island of
stability", where the expected life times of even neutron-richer isotopes
should be much longer \cite{mosel-greiner}. Indeed, life times of newly
synthesized heavy elements with Z = 108--118, increase with increasing
neutron content \cite{review3}. This is consistent with theoretical
predictions of the island of stability at the neutron number
N$\approx$184 and proton numbers around Z=114, Z=120, and eventually also
Z=126. Even more neutron-rich nuclei could be produced in multi-nucleon
transfer processes in low-energy collisions of heavy actinide nuclei like
U+Cm \cite{review4}. However, the cross sections of such processes are
rather low.

In this paper we discuss another possibility for synthesizing heavy and 
superheavy 
elements which should be more appropriate for obtaining neutron-rich nuclei 
up to their drip-lines. Namely, we propose to use intensive neutron fluxes 
generated by nuclear explosions. Actually, this method was already partially 
employed alongside with nuclear reactors in the 60-s \cite{Seaborg,Krivoch}. 
However, because of some technical and political constraints it was abandoned 
later. Nevertheless, this method provides the highest neutron densities 
which are not possible to reach with other terrestrial techniques. 

Our general ideas are illustrated in Fig.~1. It shows the upper end of 
the nuclear chart with the hypothetical ``island of stability''. The typical 
fusion reactions like $^{238}$U + $^{48}$Ca lead to proton-rich compound 
nuclei which lie above the $\beta$-stability line. On the other hand, the 
multiple neutron-capture reactions in nuclear explosions may bring the 
$^{238}$U nuclei to the neutron drip-line. Then, after the neutron flux 
ceased, these nuclei undergo multiple $\beta$-decay and come close to the 
island of stability from the neutron-rich side. If a single explosion is 
insufficient, then, in principle, a properly delayed second explosion can do 
the job.  

Several underground explosions carried out by scientists at the Livermore 
and Los Alamos Laboratories were dedicated to synthesizing new elements 
\cite{Seaborg,Krivoch,Bell67}. They include Par (1964), Barbel (1964), 
Tweed (1965), 
Vulcan (1966), and Cyclamen (1966). The elements up to $^{257}$Fm 
were found in explosion debris by radiochemical methods. From published 
descriptions of these experiments one can conclude that little special 
effort was made to increase the neutron flux and irradiation time for 
target nuclei as compared with standard explosions. Also it is unlikely 
that the debris material, which was extracted for the analysis, was located 
near the central zone characterized by the most intensive 
neutron exposure \cite{Chemi}. Although the experimental equipment available 
at that time was not very sophisticated, it was sufficient to demonstrate 
the high efficiency of this method by observing the absorption of up to 20 
neutrons by the U target. 
As discussed in refs.~\cite{Bell67,Bell65} more heavy elements might have 
been produced, but they might decay via spontaneous fission avoiding 
detection with radiochemical methods. These results were discussed 
in the review article by Glenn Seaborg \cite{Seaborg} with the conclusion 
that even higher-mass nuclei may be produced in explosions, if the 
neutron exposures were increased, and special targets were used. 

As will be demonstrated below, the multiple neutron absorption by heavy 
nuclei in the course of a nuclear explosion can lead to very high mass 
numbers close to the neutron drip-line. Not only a few but macroscopic 
amounts of such neutron-rich heavy and super-heavy nuclei may be produced 
in this way. Studying these processes may lead to better 
understanding of nucleosynthesis in r-processes associated with supernova 
explosions. We believe that some open questions of cosmic nucleosynthesis 
can be tested in 
nuclear explosion experiments by using special target materials and 
detection techniques. In our estimates below we use only open publications 
and Internet resources regarding the characteristics of explosive devices. 

\vspace{0.5cm}

{\bf 2. Main physical processes contributing to explosive nucleosynthesis.}

\vspace{0.5cm}

For our calculations we adopt a very simple model of the active zone associated with a low-yield nuclear explosion \cite{Sandmeier}. We assume that it can be represented as a medium consisting of fissioning nuclei, fission products, light charged particles, neutrons, and photons. This medium is characterized by a certain temperature $T$ which has typical values of $5 \div 10$ keV at the pre-expansion stage. Then we consider the time evolution of mass and charge distributions of nuclear species by taking into account all possible reactions with surrounding particles and subsequent decays. Elastic neutron collisions with nuclei, which do not change mass and charge distributions, are not considered here.
It is assumed that after each inelastic interaction the nucleus relaxes to a new equilibrium state (compound nucleus). In particular, the ($n,\gamma$) reaction is considered as a two-step process, including the neutron absorption at the first step and the photon emission at the second step. Specifically, in our analysis we include the following processes:

\vspace{0.5cm}

{\it Absorption of particles by a nucleus.}
The neutron capture by the nucleus with mass number $A$ is characterized by 
the width 
\begin{equation} \label{eq:absneu}
\Gamma_{n}^{cap}=\hbar \langle v_{rel} \sigma_{n} \rangle \rho_{n}. 
\end{equation}
Here $v_{rel}$ is relative $n$A velocity which is close to the thermal 
neutron velocity $v_{n}=\sqrt(3T/m_{N})$ (in units $c$, $m_{N}=939$ MeV, 
mass of nucleon), $\rho_{n}$ is the density of neutrons. For the neutron 
capture cross section $\sigma_n$ we take the inelastic part of the $n$A 
cross section, i.e. $\sigma_n=\sigma_{\rm tot}-\sigma_{\rm el}$. The 
corresponding evaluated data were taken from the web site \cite{CS}. Our analysis 
shows that in the neutron energy range between 2 MeV and 10 keV these cross 
sections may change by several times. Also, they show a 
considerable odd-even effect as functions of mass. For U isotopes the changes are 
between 1.5 barn for odd isotopes and 0.3 barn for even isotopes. For Cm isotopes 
the corresponding changes are between 3 barn and 0.5 barn. For our rough estimates 
below we take $\sigma_n$=1 barn for U and 2 barn for Cm, independent 
of neutron energy and mass number of the nucleus.

At relatively low temperatures considered here the cross-sections for 
capturing free protons and light charged clusters like  $\alpha$-particles 
are negligible. However, in thermonuclear explosions the 14 MeV neutrons 
from the fusion reaction ($d+t\rightarrow$ $^4$He+$n$) may transfer energy to 
$d$ and $t$ nuclei, sufficient to induce $(d,\gamma$), $(t,\gamma$) and other 
reactions \cite{Bell65}, which we do not consider here.   

The capture of a photon by the nucleus is characterized by the width
\begin{equation} \label{eq:absfo}
\Gamma_{\gamma}^{cap}=\hbar \langle v_{\gamma}\sigma_{\gamma}(E_{\gamma}) 
\rangle \rho_{\gamma}. 
\end{equation}
Here $E_{\gamma}=T(\pi^{4}/(30\cdot 1.202)) \approx 2.7T$ is the average 
energy of photons in matter, 
$v_{\gamma} \approx c$ is the in-medium velocity of photons, and the 
density of photons is given by 
\begin{equation} \label{eq:denfo}
\rho_{\gamma}=\frac{2\cdot 1.202}{\pi^{2}}\frac{T^{3}}{(c\hbar)^{3}}. 
\end{equation}

\vspace{0.5cm}

{\it Emission of particles from the nucleus.}
The width for evaporation of a particle $j$ ($j = n, p, d, t, ^{3}He, \alpha$) 
from the  compound  nucleus $(A,Z)$ is given by:
\begin{equation} \label{eq:eva}
\Gamma_{j}=\int_{0}^{E_{AZ}^{*}-B_{j}}
\frac{\mu_{j}g_{j}^{(i)}}{\pi^{2}\hbar^{2}}\sigma_{j}(E)
\frac{\rho_{A^{'}Z^{'}}(E_{AZ}^{*}-B_{j}-E)}{\rho_{AZ}(E_{AZ}^{*})}EdE.
\end{equation}
%
Here $\rho_{AZ}$ and $\rho_{A^{'}Z^{'}}$ are the level densities of the initial 
$(A,Z)$ and final $(A^{'},Z^{'})$ compound  nuclei, $E_{AZ}^{*}$ is the excitation 
energy of the initial nucleus, $E$ is the kinetic energy of an emitted particle, 
$g_{j}=(2s_{j}+1)$ is its spin degeneracy factor, 
$\mu_{j}$ and $B_{j}$ stand, respectively, for the reduced mass and separation energy 
of the ejectile.
The cross section $\sigma_{j}(E)$ of the inverse 
reaction $(A^{'},Z^{'})+j=(A,Z)$ is calculated using the optical model 
with corresponding nucleus-nucleus potential \cite{Botvina87}. 
The evaporation 
process was simulated by the Monte Carlo method and the conservation of 
energy and momentum was strictly controlled at each emission step. 

In the particular case of photon evaporation from an excited nucleus the 
corresponding width is given by \cite{Iljinov92} 
\begin{equation} \label{eq:gam}
\Gamma_{\gamma}=\int_{0}^{E_{AZ}^{*}}
\frac{E^{2}}{\pi^{2}c^{2}\hbar^{2}}\sigma_{\gamma}(E)
\frac{\rho_{AZ}(E_{AZ}^{*}-E)}{\rho_{AZ}(E_{AZ}^{*})}dE,
\end{equation}
where $E$ is the photon energy. 
The integration is performed numerically. Within the dipole approximation 
the photo-absorption cross section is expressed as: 
\begin{equation} \label{eq:gamsec}
\sigma_{\gamma}(E)=\frac{\sigma_{0}E^{2}\Gamma^{2}_{R}}
{(E^{2}-E^{2}_{R})^{2}+\Gamma^{2}_{R}E^{2}}. 
\end{equation}
Here the empirical parameters of the 
giant dipole resonance have the values $\sigma_0=2.5A$ mb, 
$\Gamma_{R}=0.3E_{R}$, and $E_{R}=40.3/A^{1/5}$ MeV.  
In the situation which we consider here the 
photon energies are in the range from about 4 MeV to 100 keV, i.e. 
considerably smaller than the resonance energy, $E_R\approx 13.4$ MeV 
for $A=250$. Thus, the photo-absorption cross section is rather small, 
$\sigma_\gamma < 0.01\sigma_0$.

Fission is an important de-excitation channel for heavy nuclei ($A>200$), 
which can compete with particle emission. It is also 
included in the Monte-Carlo simulations. 
Following the Bohr-Wheeler statistical approach we assume that
the partial width for the fission of compound nucleus is proportional
to the level density $\rho_{sp}$ at the saddle point of the fission 
barrier \cite{SMM} :
\begin{equation} \label{eq:fis}
\Gamma_{f}=
\frac{1}{2\pi\rho_{AZ}(E_{AZ}^{*})}\int_{0}^{E_{AZ}^{*}-B_{f}}
\rho_{sp}(E_{AZ}^{*}-B_{f}-E)dE,
\end{equation}
where $B_{f}$ is the height of the fission barrier which is determined 
according to the Myers-Swiatecki prescription \cite{MS}. For $\rho_{sp}$ we 
have used approximations obtained in ref.~\cite{Iljinov92} from the 
extensive analysis of nuclear fissility and $\Gamma_{n}$/$\Gamma_{f}$ 
branching ratios. 

\vspace{0.5cm}

{\it Beta-decay.}
There are many approaches to such a complicated problem as $\beta$-decay. 
In our calculations we are mostly dealing with nuclei far from the $\beta$ 
stability region. Their masses and $\beta$-decay rates have significant 
uncertainties. On the other hand, it is well known that $\beta$-decay is 
rather slow in comparison with other nuclear processes, it has characteristic 
times of milliseconds and more. A maximum effect of $\beta$-decay was 
estimated 
by taking the rates for 'allowed' transitions from 
the Sargent diagram \cite{Sargent}. In particular, we have used the following 
interpolation of the decay rate  $\lambda$ (in 1/sec): 
\begin{equation} \label{eq:betadecay}
log(\lambda) = 4log(E_{max})-3, 
\end{equation}
where $E_{max}$ (in MeV) is the maximum energy of emitted electrons, 
assuming that the daughter nucleus is in its ground state. Then the 
corresponding decay width is $\Gamma_{\beta}=\hbar \lambda$. For nuclei in 
the vicinity of the drip-line this formula predicts life times of order 
of milliseconds \cite{Zagreb}. 

\vspace{0.5cm}

{\it Spontaneous fission and $\alpha$-decay.}
Spontaneous fission is characterized by even longer times than $\beta$-decay. 
It will certainly affect the yield of superheavy elements before 
extraction of a sample for analysis. However, it has practically no 
influence on the accumulation of neutrons by the target nucleus. To take into 
account this process we have employed a simple parametrization of the 
fission half-lifes proposed in ref.~\cite{Bell67}. Typical half-lives for 
alpha-decay of these neutron-rich nuclei are usually longer than years and this 
decay mode can be completely ignored in the calculations.


\vspace{0.5cm}

{\bf 3. Physical conditions in the active zone}

\vspace{0.5cm}

{\it Neutron density:} 
For our further calculations we should know the density of neutrons 
generated in the active zone of an explosion. Let us assume a 100 kt 
explosion which corresponds to fissioning of 1.45$\cdot$10$^{25}$ nuclei 
of $^{235}$U (or $^{239}$Pu) with total mass of 5.6 kg. 
At normal density this amount of U or Pu has the volume of about 300 cm$^{3}$. 
Assuming that about 1.3 new neutrons are produced in average per fission, 
we can estimate the total number of free neutrons to be  2$\cdot$10$^{25}$. 
This gives the maximum neutron density of about 6$\cdot$10$^{22}$ cm$^{-3}$. 
However, due to a finite efficiency, the amount of the fissile material and 
accordingly its volume should be larger. On the other hand, with the implosion 
assembly method, the fissile material is initially compressed 
by several times, that will increase accordingly the neutron density.  
According to ref.~\cite{Sandmeier}, the neutron multiplication 
process should be over within the time interval 
of about 1.5$\cdot$10$^{-7}$ sec after beginning of the chain reaction. 
To this time the density of free neutrons increases exponentially and 
reaches the maximum values estimated above. Many of these fission neutrons 
may leave the core. However, some of them might be reflected back with an 
appropriate reflector to keep the neutron density in the core higher. 
These neutrons have a chance to slow down to thermal energies corresponding 
to temperatures $T \approx 10$ keV at the early expansion stage. 
Taking into account all these uncertainties as well as strong time dependence 
of the neutron yield, in our estimates below we use several values for the 
neutron density, from $4\cdot 10^{22}$ cm$^{-3}$ to $4\cdot 10^{20}$ cm$^{-3}$.   

One can also consider a thermonuclear reaction as a source of additional fast
neutrons in combined fission-fusion set-ups. We have found that the 
neutron densities at the end of the fusion stage are rather similar to the 
values estimated above. However, the subsequent evolution of the neutron 
density will be different in these two cases, because of a very different 
composition of the active zone. 

{\it Characteristic time scales:} There are several important 
time scales which influence strongly the efficiency of the neutron capture 
reaction. Obviously, the goal is to maximize the time $\tau$ of the target exposure to 
the strongest neutron flux. The shortest time scale is the neutron multiplication 
time $\tau_{m}$ associated with the chain reaction. This time is determined by 
the properties of the fissile material and is typically in the range of 
0.1 $\mu$sec, as mentioned above. Such short times are obtained when all 
available neutrons are used for fission reactions, to maximize the yield of 
the explosion. However, this time will increase significantly if the active 
zone would contain more nuclei absorbing neutrons, like $^{238}$U (reduced 
enrichment), bringing the system closer to the critical regime. 

The disassembly stage is defined by a condition that the chain reaction is ceased. 
This happens when most of the fission energy is deposited in the active zone and 
the matter is transformed 
into a radiation-dominated hot and dense plasma. Due to the huge internal pressure 
it would rapidly expand into the vacuum, if nothing is done to confine it. However, 
by introducing a heavy tamper around the active zone, one can delay the fast expansion 
of plasma to a few microseconds. This time is needed for the shock wave to 
propagate through the thick layer of heavy tamper's material, usually $^{238}$U.
The hydrodynamic time can be estimated as 
a time required for the rarefaction wave to reach the center of the active zone,
$\tau_{h} \approx R/c_s\loo 10 \mu$sec, where $R$ is the core radius and 
$c_s$ is the sound velocity in the core. 
The tamper can also serve as a reflector and absorber for the neutrons. 
To explore the whole parameter space, in our calculations we consider the 
exposure times $\tau$ from 0.1 to 3 $\mu$sec. 

Another important time scale is the neutron thermalization time $\tau_{th}$ 
in the medium. If the medium consists mostly of heavy nuclei like U, the 
neutron energy loss is dominated by inelastic collisions with these 
nuclei, followed by the photon emission from the excited nuclei. For slowing 
down the fission neutrons to thermal velocities $v_{n}(T)$ 
several ($\sim 10$) such collisions are needed. Therefore, we can estimate 
thermalization time as $\tau_{th} \sim 10\lambda/v_{n}$, where $\lambda$ 
is the neutron mean free path (few cm). 
Then one obtains $\tau_{th} \sim 1 \mu$sec, 
which is usually shorter than the hydrodynamic time. 

\vspace{0.5cm}

{\bf 4. Synthesis of nuclei by multiple neutron capture}

\vspace{0.5cm}

{\it Evolution of average mass number:} 
We have performed calculations for the $^{238}$U and $^{248}$Cm target 
nuclei embedded in the active zone of the explosion for several fixed values 
 of neutron density $\rho_n$ and temperature $T$. The $^{238}$U nuclei are 
usually present in the explosive devices in the fissile material or 
in a tamper, but  $^{248}$Cm nuclei can be additionally installed in the 
core.\footnote{The Cm isotopes 
are produced in significant quantities in nuclear reactors, and they have 
a sufficient lifetime to be used as a target for the neutron irradiation.} 
Instead of solving a complicated set of rate equations for the ensemble of different nuclear spices, we first consider the fate of one individual target nucleus in this environment. We start at $t$=0 with a nucleus $A$ in the ground state and consider all processes which can change its mass number or excitation energy. At each time step we evaluate the probabilities of these processes, $p_j=\Gamma_j/\left(\sum_k\Gamma_k\right)$, in accordance with the rates $\Gamma_j$ discussed in the previous section. Then we choose the reaction by the Monte Carlo method and change accordingly the state of the nucleus. If process $j$ is chosen, the clock is changed by time interval $\Delta t=\hbar/\Gamma_j$. This method allows us to follow the stochastic evolution of an "average" nucleus.

The average mass numbers of U and Cm isotopes are shown in Fig.~2 as functions of time. 
The dotted lines show the calculations for fast fission neutrons 
(mean kinetic energy of 1.5 MeV) assuming no moderation at all. 
The other lines show the results for thermalized 
neutrons at temperature $T$=10 keV and several neutron densities. 
Due to a higher flux, the fast neutrons dominate at early times, $t<0.3$ $\mu$sec.
However during this time interval the target nuclei can absorb only a few neutrons.  
The efficient nucleosynthesis may proceed only at longer times, when the 
neutrons have already slowed down. Then, due to the lower excitation energy, 
the probability of channels with re-emission of neutrons from the compound 
nucleus is strongly reduced as compared with reactions induced by MeV-neutrons. 
This is why with inceasing $A$ the absorption rate becomes higher for 
keV-neutrons. The calculations show that a considerable 
part of curium (more than 90\%) undergoes fission in reactions with fast 
neutrons. However, in the case of thermalized neutrons this channel
is less probable and the majority of nuclei (nearly 95\%) survive. 
One can see from the figure that U and Cm nuclei can in average absorb 10-15
neutrons during the time interval of 1 $\mu$sec, if the neutron density 
remains high, 4$\cdot$10$^{22}$ cm$^{-3}$ or higher. Since the neutron 
capture process has a stochastic character, even much larger numbers of neutrons 
could be accumulated with certain probabilities (see the next subsection). 

The main mechanism of 
increasing the nuclear mass number works as following. A neutron 
captured by a nucleus with mass $A$ brings a few MeV excitation to the  
new nucleus $A$+1. This excitation energy is only slightly above the neuteron 
separation energy, which is expected to decrease with the neutron excess. It is 
typically below the fission barrier, especially in neutron-rich isotopes. 
The probability to return a neutron in the evaporation process is also low 
because of the small phase space volume of the final state. In this situation  
$\gamma$-emission is the main decay mechanism for reducing the 
excitation energy. After the first photon carries away some energy, the next 
$\gamma$-emissions become even more probable in comparison with other decay 
channels. This continues until the excitation is decreased nearly to zero, 
and then the relative probability to capture a new neutron becomes high 
again. With fixed $Z$ and increasing $A$ the fission probability decreases, 
since the fissility parameter $Z^{2}/A$ becomes lower. In the time interval 
$\loo$1 msec the $\beta$-decay still has a very low probability 
in order to compete with other processes. 
In the long run the survival of produced superheavy nuclei will 
depend on whether they can avoid spontaneous fission and $\alpha$-decay in 
order to reach the island of stability via multiple $\beta$-decays. 

As our calculations show, the outcome of explosive nucleosynthesis depends 
crucially on the masses and  level densities of nuclei far away from the 
$\beta$-stability line. There are no experimental data in this region, and theoretical 
models often give controversial results. Our previous calculations were performed 
with Myers-Swiatecki (MS) mass formula \cite{MS}. The predictions based on the 
Cameron mass formula \cite{Cameron} are shown in Fig.~3 for U and Cm targets 
too. In principal, if the neutron flux remains high the capture of neutrons 
may continue until the target nucleus reaches the neutron drip-line. Then 
the probability of neutron emission increases considerably, and, in the case 
of U, it dominates over the photon emission. The ``dynamical'' equilibrium 
between absorption and emission of neutrons at the drip-line depends weakly 
on the neutron density and may last for a long time (more than milliseconds) 
until the $\beta$-decay makes the $Z$+1 nucleus. As the result, the drip-line 
is shifted to the right and absorption of neutrons becomes possible again, 
of course, if the neutron density remains high enough. One can see that 
the Cameron formula leads to similar results for U and Cm nuclei, with 
saturation at large  $A$. However, the MS formula predicts slightly different 
Cm masses around $A \approx 280$ as compared to the Cameron formula. 
This leads to a smaller ratio of the neutron and photon evaporation widths, 
$\Gamma_{n}/\Gamma_{\gamma}$, so that the emission of photons is more probable 
in the MS case. This gives a chance to absorb more neutrons 
in Cm target, contrary to the U target, and, finally, enter the mass region 
with more bound neutrons. 

\vspace{0.5cm}

{\it Calculation of nuclear abundances:} 
For practical applications, it is very important to estimate the amount 
of new isotopes produced by this mechanism. In Figs.~2 and 3 we have 
presented the evolution of an average mass number of the target nucleus, ignoring 
the distribution around the average. This analysis has provided us 
with better understanding of the main mechanisms responsible for increasing the 
nuclear mass number under various external conditions. As turned out, at 
temperatures $T \sim 10$ keV we can simply consider a capture of neutrons 
followed by the emission of photons, without complications induced by 
re-emission of neutrons and fission. 
In this case we can write the set of coupled equations for densities of 
nuclei with mass numbers $A$=$A_0$, $A_0$+1, $A_0$+2, ... :
\begin{eqnarray} \label{eq:den}
\frac{d \rho_{A_0}}{d t}=-\langle v_{rel} \sigma_{n} \rangle \rho_{n} 
\rho_{A_0},
\nonumber \\
\frac{d \rho_{A+1}}{d t}=\langle v_{rel} \sigma_{n} \rangle \rho_{n} \rho_{A} 
-\langle v_{rel} \sigma_{n} \rangle \rho_{n} \rho_{A+1}, 
\end{eqnarray}
which can be solved numerically for any given time 
dependence of the neutron density $\rho_{n}$. 
Here $v_{rel}$ is the neutron-nucleus relative velocity which is close to 
the neutron velocity $v_{n}$, and $\sigma_{n}$ is the neutron capture 
cross-section (see section {\bf 2}), which is assumed here to be independent 
of the nuclide mass number.\footnote{ 
Of course, when the neutron number approaches the drip-line limit, 
the capture cross section $\sigma_{n}$ will strongly decrease, 
and the solution of the equations will be more complicated.} 
These equations describe the evolution of the ensemble of nuclear 
species due to the neutron capture reactions with gain ($A-1 \rightarrow A$) 
and loss ($A \rightarrow A+1$) terms. It is worth noting that final nuclide 
abundances depend only on the time-integrated neutron flux 
$\Phi = \int \rho_{n}v_{n} dt\approx v_n\rho_n\tau$. 
According to the analysis of refs.~\cite{Krivoch,Bell67,Bell65}, typical 
values of $\Phi$ reached in low-yield nuclear explosions are in the range of 
3$\cdot10^{24} \div 10^{25}$ $n$/cm$^{2}$.  

In Fig.~4 we present the 
distributions of isotope abundances after exposure of a $^{238}$U target to 
a constant neutron flux during fixed time intervals indicated at the lines. 
The figure shows that the distribution of neutron-rich isotopes becomes 
broader with increasing the exposure time, or, equivalently, the integrated 
neutron flux $\Phi$. According to refs.~\cite{Krivoch,Bell65}, in samples of 
material extracted from underground explosion sites concentrations of rare nuclei 
(A=257) on the level of $10^{-11} - 10^{-12}$ were measured by radiochemical methods. 
Even higher sensitivity can be reached by using mass spectroscopic methods 
(see, e.g., \cite{Dellinger}). In typical nuclear explosions mentioned above 
the observed neutron-rich isotopes have not more than 20 absorbed neutrons
~\cite{Krivoch,Bell65}. According to calculations of Fig.~4, where neutron 
density was fixed at 4$\cdot$10$^{22}$ cm$^{-3}$, to produce such isotopes in 
observable amounts we would need $\tau \approx 0.3 \mu$sec. This corresponds to the 
integrated neutron flux of about $2\cdot 10^{24}$ $n$/cm$^{2}$, which within a 
factor of 2  agrees with the values estimated in refs.~\cite{Krivoch,Bell65}. 

From Fig.~4 we conclude that observable concentrations of nuclei with more 
than 40 absorbed neutrons can be obtained only if the exposure time is longer than 
1$\mu$sec. Obviously, to increase the number of captured neutrons one should 
increase the integrated neutron flux $\Phi$, i.e. 
either by increasing the neutron density $\rho_n$ or by increasing the exposure 
time $\tau$.\footnote{We are taking here about time scales which are still much 
less than the $\beta$-deacy times, i.e. milliseconds.} 
To achieve this goal one can try e.g. to increase 
the neutron multiplication time by introducing neutron absorbers, 
reflectors, and other construction elements. With the exposure time of 
3 $\mu$sec the fraction of nuclei captured 50 neutrons would be about 
10$^{-8}$. In this case, even if we take 1 gram of the exotic target 
material like $^{248}$Cm, the concentration of nuclides with A$\approx$300 
will be  10$^{-12}$ which could be detected by present experimental methods. 

These results suggest that the drip-line region can in principle be 
reached in a properly optimized explosive processes. Obviosly, new 
experiments are required to explore this region. They should not only 
be limited to searching for long-lived superheavy elements, 
where the yield may be essentially affected by spontaneous fission. 
Heavy and intermediate-mass nuclei in the vicinity of their neutron drip-lines 
could be synthesized in this way too. For example, we have found that under 
the same conditions a Pb nucleus may capture nearly 40 neutrons and come 
close to the drip-line. This approach to the drip-line has an important 
advantage as compared with heavy-ion reactions, because nuclei enter this 
region with minimal excitation energy. However, due to expected short live-times of 
these nuclei, their properties can only be studied by analizing their decay products.

\vspace{0.5cm}

{\bf 5. Conclusions}

\vspace{0.5cm}

It is obvious that the nuclear explosions provide the highest fluxes of 
neutrons under terrestrial conditions which will hardly be feasible with 
any other experimental technique. 
We have demonstrated that irradiation of nuclei with neutrons produced 
during the first microseconds of a nuclear explosion is a very promising 
way to synthesize neutron-rich heavy and super-heavy elements. For comparison, 
modern spallation sources (e.g. ESS) may supply thermal neutrons with 
densities around 
10$^{12}$ cm$^{-3}$ , and nuclear reactors -- with 10$^{11}$ cm$^{-3}$ 
\cite{neutron-sources}. 
We believe, there exist technical possibilities to increase the time and 
intensity of target exposure in the course of a nuclear explosion. 
They may include preliminary compression of fissile material, introducing 
neutron reflectors and moderators, as well as special construction of 
targets. The main idea of optimization should be to maximize the neutron 
density and exposure time and to minimize the energy yield of the explosion. 
Another possibility which should be considered in more details is 
generating two or several nuclear explosions within a time delay 
up to milliseconds in close proximity of each other. 
This delay may be sufficient for very neutron-rich nuclei 
produced in the first explosion to undergo multiple $\beta$-decay. 
Then the daughter nuclei will be able to absorb additional 
neutrons from the second explosion, and in this way increase the resulting 
mass number, as illustrated in Fig.~1. Even mass production 
of super-heavy elements can be envisaged in the future. 

The methods discussed above can be used not only for synthesizing long-living 
superheavy elements but also for production of very neutron-rich isotopes in 
the vicinity of the drip-line. These studies are extremely important for 
understanding the origin of heavy elements in the Universe. Hot and dense 
environments created in nuclear explosions are rather similar to conditions 
in supernova explosions, when heavy nuclei are produced by multiple neutron 
capture in $r$-process. 

If long-lived superheavy elements are indeed produced in the explosions, an 
important question to be answered is, how to find them among tons of debris. 
In underground explosions the processed material may be extracted and 
analysed by radio-chemical and mass-spectroscopic methods. Generally, the 
methods used for searching for superheavy elements in nature \cite{nature} 
may be applied in this case too.

The authors thank Yu. Ts. Oganessian, I. Pshenichnov and S. Schramm for 
fruitful discussions. A.B. acknowledges the financial support received from 
the Helmholtz International Center for FAIR within theframework of the LOEWE 
program (Landesoffensive zur Entwicklung Wissenschaftlich-Okonomischer Exzellenz)
launched by the State of Hesse. This work was supported in part by the grants 
NS-7235.2010.2 and RFBR 09-02-91331 (Russia).

\vspace{2cm}

\begin{figure} [tbh]
\begin{center}

\vspace{-1cm}

\hspace{2cm}

\includegraphics[width=15cm]{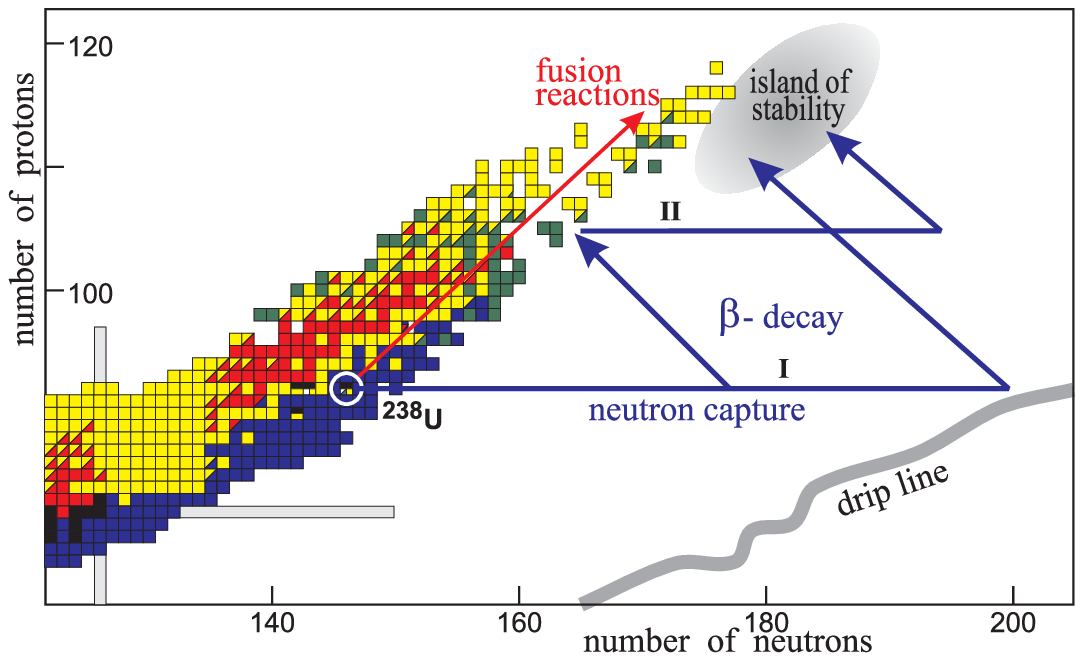}

\caption{\small{(Colour online) 
The upper end of the nuclear chart showing the 
experimentally investigated isotopes. The colours stand for 
the decay modes: yellow - $\alpha$-decay, red - $\beta^{+}$ or electron 
capture, blue - $\beta^{-}$ decay, green - spontaneous fission. Black 
boxes are stable nuclei. Grey area is the predicted ``island of stability''. 
Red arrow represents the heavy-ion  fusion reactions used so far.
Blue arrows show schematically the proposed explosion-based methods to reach 
this island: I indicates fast capture of neutrons during first microseconds 
of the nuclear explosion followed by the $\beta$-decay. II indicates second 
nuclear explosion within few milliseconds after the first one. 
}}

\end{center}
\end{figure} 

\begin{figure} [tbh]
\begin{center}

\vspace{-1cm}

\hspace{1cm}
\includegraphics[width=14cm]{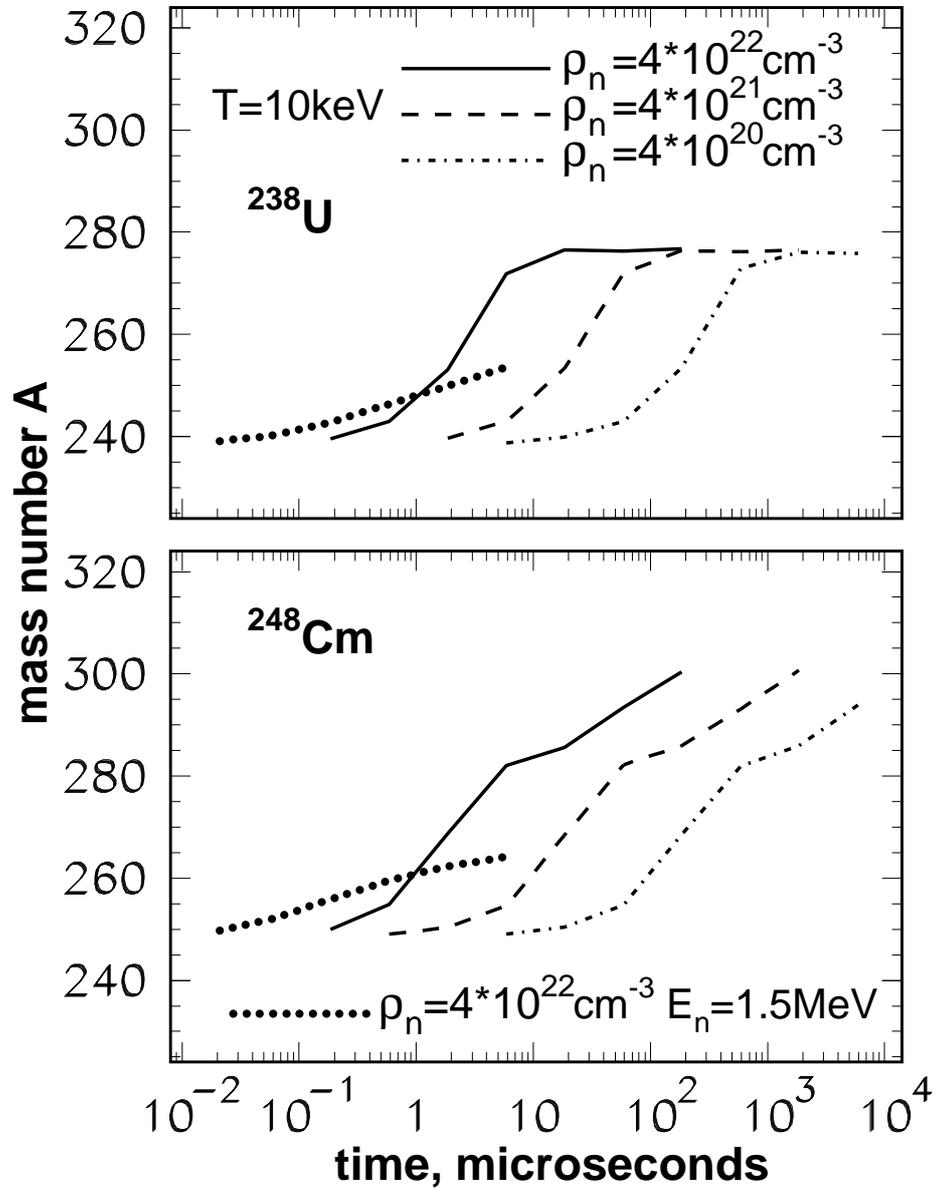}

\caption{\small{Evolution of U (top panel) and Cm (bottom panel) mass 
numbers with time at the neutron densities shown at the figure. (The 
initial target nuclei are $^{238}$U and $^{248}$Cm.) Dotted 
line presents absorption of fast neutrons, other lines - absorption of 
thermalized neutrons at T=10 keV. }}

\end{center}
\end{figure} 

\begin{figure} [tbh]
\begin{center}

\vspace{-1cm}

\hspace{2cm}

\includegraphics[width=14cm]{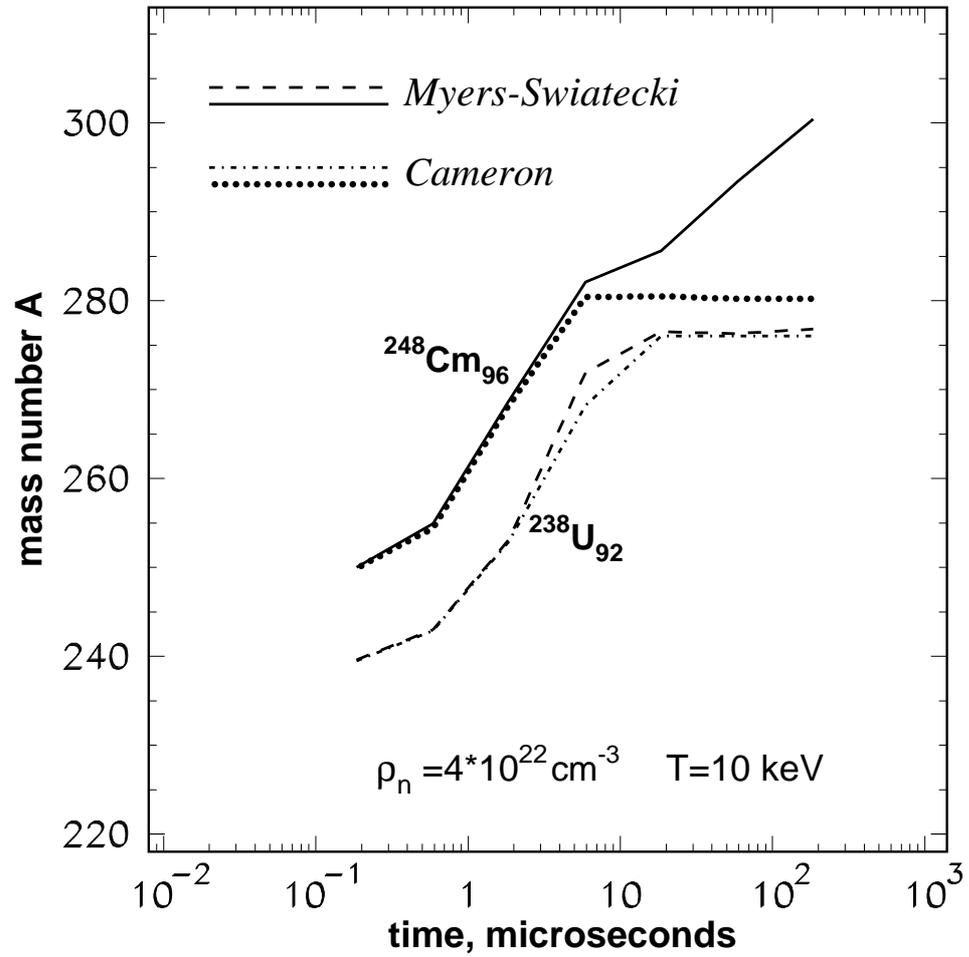}

\caption{\small{Effect of different mass formulae on calculations of mass 
number evolution for $^{238}$U and $^{248}$Cm targets. Dashed and solid 
lines: masses are taken from ref.~\cite{MS}, dotted and dash-dotted lines --
ref.~\cite{Cameron}. Neutron density, temperature are also shown in the 
figure. }}

\end{center}
\end{figure} 

\begin{figure} [tbh]
\begin{center}

\vspace{-1cm}

\hspace{2cm}

\includegraphics[width=16cm]{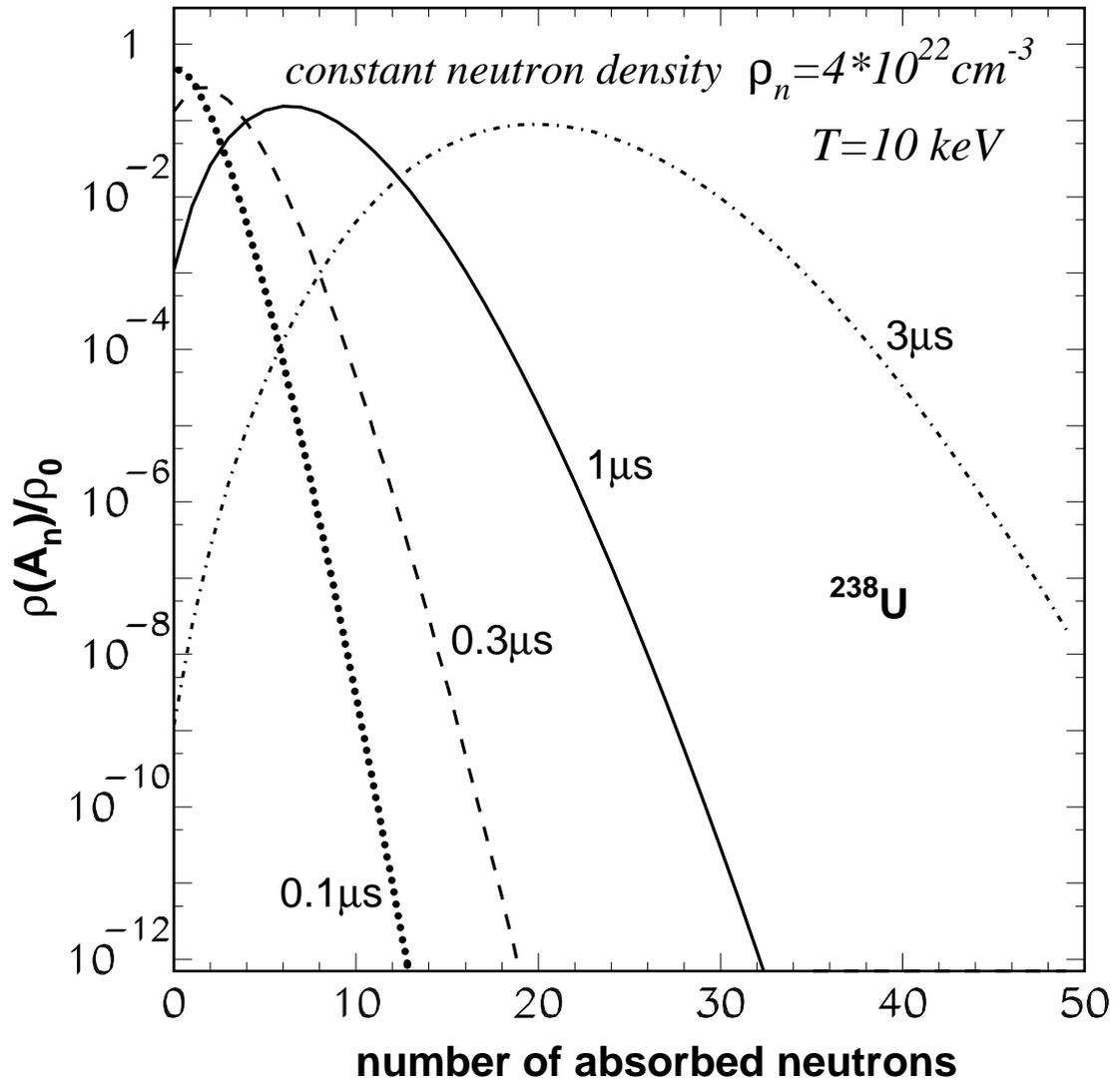}

\caption{\small{Densities of nuclei produced after capture of neutrons 
normalized to the initial density of $^{238}$U target. Density and 
temperature of neutron, as well as exposure times (in microseconds) 
are shown in the figure. }}

\end{center}
\end{figure} 

\end{document}